\documentclass[aps,twocolumn,PRL,superscriptaddress,reprint]{revtex4-1}
\usepackage{amsmath}
\usepackage{epsfig}
\usepackage{graphicx}
\usepackage{graphicx, color, epstopdf}
\usepackage{bm}
\usepackage{amssymb}
\usepackage{hyperref}
\usepackage{xcolor}
\usepackage{subfigure}
\usepackage{makecell}
\usepackage{tikz}

\usepackage{etoolbox}
\usepackage[version=4]{mhchem}

\usepackage{xcolor}
\usepackage{etoolbox}

\makeatletter
\newcommand{\mycolorhook}[1]{%
  \edef\tmpID{\detokenize{#1}}%
}

\let\old@bibitem\@bibitem
\def\@bibitem#1{%
  \mycolorhook{#1}
  \old@bibitem{#1}
}

\let\old@lbibitem\@lbibitem
\def\@lbibitem[#1]#2{%
  \mycolorhook{#2}
  \old@lbibitem[#1]{#2}
}

\patchcmd{\NAT@bibitem@first@sw}{\NAT@anchor{#1}}{\mycolorhook{#1}\NAT@anchor{#1}}{}{}
\makeatother

\preto\section{%
  \ifnum\value{section}=4 %
  \fi
}

\hypersetup{hidelinks,
	colorlinks=true,
	allcolors=black,
	pdfstartview=Fit,
	breaklinks=true}
	
\definecolor{lime}{HTML}{A6CE39}
\DeclareRobustCommand{\orcidicon}{
\begin{tikzpicture}
\draw[lime, fill=lime] (0,0)
circle[radius=0.16]
node[white]{{\fontfamily{qag}\selectfont \tiny \.{I}D}};
\end{tikzpicture}
\hspace{-2mm}
}
\foreach \x in {A, ..., Z}{%
\expandafter\xdef\csname orcid\x\endcsname{\noexpand\href{https://orcid.org/\csname orcidauthor\x\endcsname}{\noexpand\orcidicon}}
}

\begin{document}
\bibliographystyle{apsrev4-1}
\title{{Altermagnetic Spin Precession and Spin Transistor}}
\affiliation{National Laboratory of Solid State Microstructures, School of Physics,
and Collaborative Innovation Center of Advanced Microstructures,
Nanjing University, Nanjing 210093, China} 
\affiliation{Jiangsu Physical Science Research Center and Jiangsu Key Laboratory of Quantum Information Science and Technology, Nanjing University, Nanjing 210093, China} 
\affiliation{School of Physics and Astronomy, Yunnan Key Laboratory for Quantum Information, Yunnan University, Kunming 650091, People’s Republic of China} 

\author{Li-Shuo Liu}
\affiliation{National Laboratory of Solid State Microstructures, School of Physics,
and Collaborative Innovation Center of Advanced Microstructures,
Nanjing University, Nanjing 210093, China}

\author{Kai Shao}
\affiliation{National Laboratory of Solid State Microstructures, School of Physics,
and Collaborative Innovation Center of Advanced Microstructures,
Nanjing University, Nanjing 210093, China}
\affiliation{School of Physics and Astronomy, Yunnan Key Laboratory for Quantum Information, Yunnan University, Kunming 650091, People’s Republic of China}

\author{Hai-Dong Li}
\affiliation{National Laboratory of Solid State Microstructures, School of Physics,
and Collaborative Innovation Center of Advanced Microstructures,
Nanjing University, Nanjing 210093, China}

\author{Xiangang Wan}
\affiliation{National Laboratory of Solid State Microstructures, School of Physics,
and Collaborative Innovation Center of Advanced Microstructures,
Nanjing University, Nanjing 210093, China}

\author{Wei Chen \hspace{-1.5mm}\orcidA{}}
\email{Corresponding author: pchenweis@gmail.com}
\affiliation{National Laboratory of Solid State Microstructures, School of Physics,
and Collaborative Innovation Center of Advanced Microstructures,
Nanjing University, Nanjing 210093, China}
\affiliation{Jiangsu Physical Science Research Center and Jiangsu Key Laboratory of Quantum Information Science and Technology, Nanjing University, Nanjing 210093, China}

\author{D. Y. Xing}
\affiliation{National Laboratory of Solid State Microstructures, School of Physics,
and Collaborative Innovation Center of Advanced Microstructures,
Nanjing University, Nanjing 210093, China}

\begin{abstract}
Altermagnets hold great potential for spintronic applications, yet their intrinsic spin dynamics and associated transport properties remain largely unexplored. Here, we investigate spin-resolved quantum transport in a multi-terminal setup based on a $d$-wave altermagnet. It is found that the altermagnetic spin splitting in momentum space induces an interesting spin precession in two-dimensional real space, giving rise to characteristic spin patterns. This altermagnetic spin precession manifests as a spatial modulation of the transverse Hall-like voltage, whose oscillation period provides a direct measure of the spin-splitting strength. When the altermagnetism is electrically tunable, the proposed setup functions as a prototype for a highly efficient spin transistor. The key physical effects are shown to be robust against dephasing and crystalline warping. Our work not only identifies a fingerprint signature of altermagnets, offering a direct probe of the altermagnetic spin splitting, but also represents an important step toward bridging their fundamental physics with practical spintronic applications.

\end{abstract}

\date{\today}

\maketitle

\emph{Introduction.---}Various magnetic phases are conventionally classified based on the arrangement of magnetic atoms, with the role of nonmagnetic constituents typically neglected~\cite{neel1971magnetism,vsmejkal2022emerging}. However, recent studies on altermagnetism have revealed that nonmagnetic atoms can play a significant role in shaping the magnetic properties of the system~\cite{vsmejkal2022emerging}. In particular, the antiferromagnetic sublattices with opposite spins cannot be related by inversion or translation operations~\cite{vsmejkal2022beyond}, giving rise to pronounced momentum-dependent spin splitting in the band structure, despite the absence of net magnetization~\cite{wu2007fermi,hayami2020bottom,naka2021perovskite,yuan2021prediction,yuan2020giant,naka2019spin,hayami2019momentum,ahn2019antiferromagnetism,vsmejkal2022anomalous,ma2021multifunctional,mazin2021prediction}. Notably, the spin splitting in altermagnets can exceed that induced by spin-orbit coupling by an order of magnitude and exhibits unconventional $d$-, $g$-, and $i$-wave symmetries~\cite{vsmejkal2022beyond,krempasky2024altermagnetic,lin2024observation}, opening up promising opportunities for novel spintronic applications.

\begin{figure}[!htbp]
    \centering
    \includegraphics[width=0.48\textwidth]{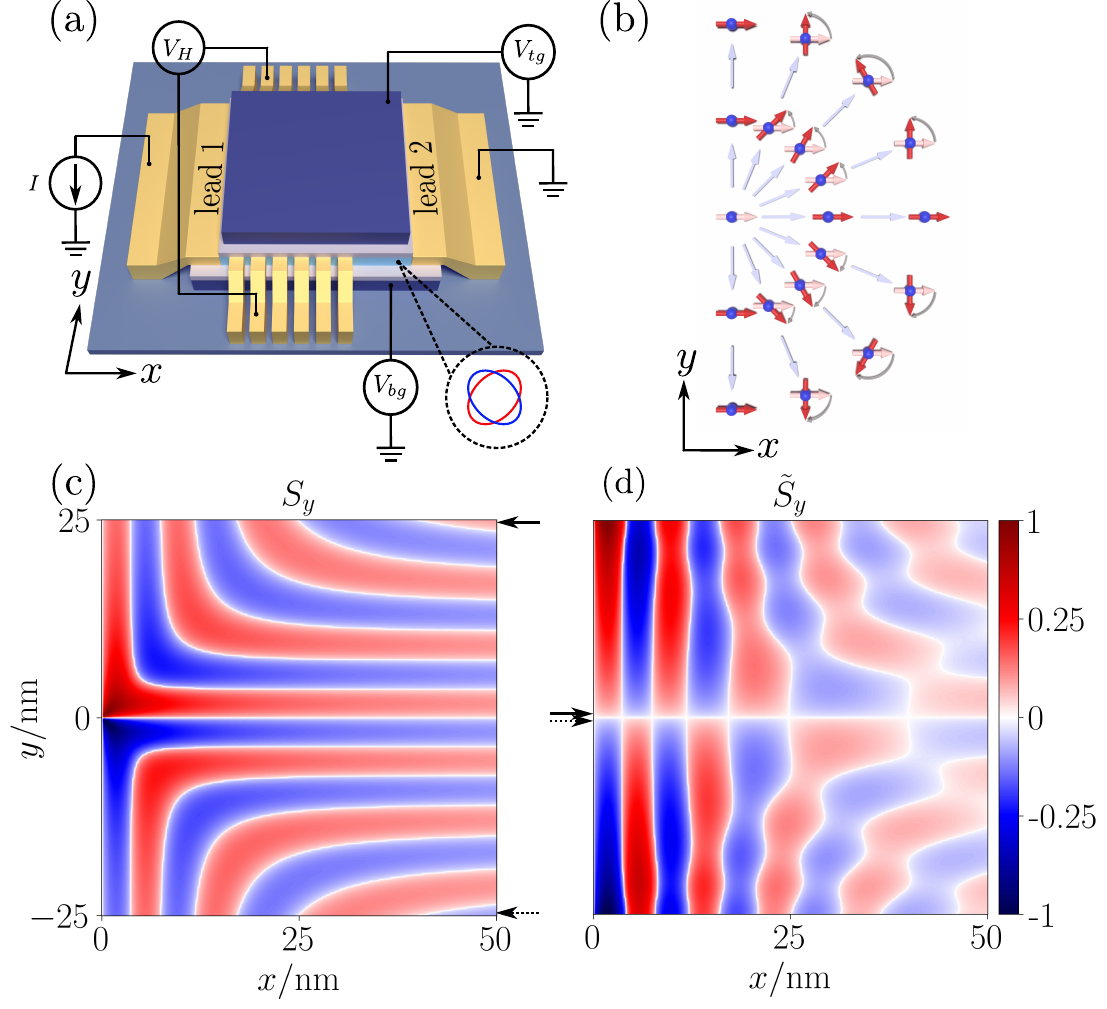}
    \caption{
   (a) Schematic of the proposed setup, consisting of a central altermagnetic region connected to two longitudinal leads (1 and 2) and multiple transverse voltage probes. Top and bottom gates ($V_{tg}$ and $V_{bg}$) control the altermagnetic splitting, which is essential for the spin transistor. The $d$-wave altermagnetic Fermi surfaces are illustrated in the inset. 
   (b) Altermagnetic spin precession in real space, with arrows indicating the local spin orientations.
(c,d) Spatial distribution of spin component $S_y$ for nonequilibrium propagating electrons under (c) point injection and (d) line injection, with the injected spins polarized along the $x$ direction. The solid and dashed arrows highlight the corresponding spin modulation along the $x$ direction in (c) and (d). A square-root scale is applied to the colorbar to enhance the contrast around zero.
Relevant parameters: $B = 0.2 \mathrm{~eV}\cdot \mathrm{nm}^2,~\alpha_{A} = 0.08\mathrm{~eV}\cdot \mathrm{nm}^2,~\mu = 0.1\mathrm{~eV}$.
}
    \label{fig1}
\end{figure}

Significant progress has been made in the search for altermagnetic materials, both theoretically and experimentally~\cite{bai2024altermagnetism, song2025altermagnets}. 
In particular, first-principles calculations have predicted hundreds of potential altermagnetic candidates~\cite{bai2024altermagnetism, sodequist2024two, vsmejkal2022beyond, vsmejkal2022emerging, jiang2025metallic, zhang2025crystal, hu2025catalog, chen2024enumeration}, and altermagnetic spin splitting has been experimentally observed through angle-resolved photoemission spectroscopy~\cite{zhu2024observation,fedchenko2024observation, krempasky2024altermagnetic, ding2024large, lin2024observation, reimers2024direct, jiang2025metallic, zhang2025crystal, yang2025three, jungwirth2025symmetry}.
A more refined characterization of the altermagnetic Fermi surface, through quantum oscillation measurements~\cite{li2025diagnosing} and quasiparticle interference imaging~\cite{hu2025quasiparticle,lou24prb,linder24prb}, is warranted. 
The distinctive band structures of altermagnets can lead to a range of notable physical phenomena, including the anomalous Hall effect~\cite{vsmejkal2020crystal,feng2022anomalous,vsmejkal2022anomalous,reichlova2024observation,gonzalez2023spontaneous,zhou2025manipulation, galindez2025revealing, leiviska2024anisotropy, li2025exploration,yu2025neel}, giant magnetoresistance~\cite{vsmejkal2022beyond,vsmejkal2022giant,li2025marginal}, magnetic nonlinear Hall effect~\cite{han2025discovery}, spin-charge conversion~\cite{karube2022observation,bai2023efficient,gonzalez2021efficient,jechumtal2025spin}, interesting Josephson effects~\cite{li2025spin,zhang2024finite}, novel magnetoelectric coupling~\cite{chen2025electrical,duan2025antiferroelectric,gu2025ferroelectric} and Hall drag effects~\cite{lin2025coulomb}, 
and other interesting transport properties~\cite{vakili2025spin, kokkeler2025quantum,zhang2025electrical,zarzuela2025transport}.
While these advances highlight the great potential of altermagnets for spintronic applications~\cite{jungwirth2025altermagnetic}, their characteristic spin dynamics and the resulting quantum transport properties in multi-terminal measurements remain poorly understood.

In this work, we address this gap by investigating spin-resolved quantum transport in a multi-terminal device based on a two-dimensional $d$-wave altermagnet, as illustrated in Fig.~\ref{fig1}(a).
Owing to the altermagnetic spin splitting, electrons injected with spin polarization perpendicular to the N\'eel vector of the altermagnet undergo a characteristic spin precession [Fig.~\ref{fig1}(b)], resulting in distinctive real-space spin patterns [Figs.~\ref{fig1}(c,d)]. These spin patterns faithfully reflect the underlying altermagnetic spin splitting in momentum space and can be probed through the spatial oscillations of the transverse Hall-like voltage. The oscillation period, in particular, is an analytical function of the altermagnetic spin splitting, providing a direct measure of this key physical parameter. The physical effects are robust against both dephasing and crystalline warping. Moreover, for electrically tunable altermagnets, the proposed setup naturally functions as a prototype for a high-efficiency spin transistor. Our results establish a fingerprint signature of altermagnets and pave the way for their device applications.

\emph{Altermagnetic spin precession.---}We consider a two-dimensional $d$-wave altermagnet, specifically with $d_{xy}$-type order, to illustrate the unique spin precession mechanism.
The system is modeled by the effective Hamiltonian as~\cite{vsmejkal2022beyond}
\begin{equation}\label{kp}
    H = \frac{B}{2}\left(k_x^2 + k_y^2\right) + \alpha_{A} k_x k_y \sigma_z - \mu,
\end{equation}
where $B$ denotes the inverse effective mass of the electron, $\alpha_{A}$ characterizes the altermagnetic spin-splitting, $\sigma_{z}$ (and later $\sigma_{x,y}$) are the Pauli matrices representing electron spin, and $\mu$ is the chemical potential.

The altermagnetic spin splitting can be viewed as a momentum-resolved Zeeman field, leading to distinctive spin precession and, consequently, characteristic spin patterns in real space.
To illustrate this, we first consider the local injection of electrons at the origin without loss of generality.
The propagator from the origin to $\bm{r}$ takes the form~\cite{SM}
\begin{equation}
    g^R_{\sigma,\sigma'}(\bm{r};\omega) = - \frac{4\pi\delta_{\sigma,\sigma'}}{\sqrt{B^2- \alpha_{A}^2}}K_0(-i {\bm{k}}_{\sigma}\cdot\bm{r}),
\label{eq2}
\end{equation}
where the subscripts $\sigma,\sigma'=\uparrow,\downarrow$ label the spin components and $K_0(\cdot )$ is the zeroth-order modified Bessel function of the second kind. The Kronecker delta $\delta_{\sigma, \sigma'}$ indicates the absence of spin flipping during electron propagation, a consequence of the collinear spin splitting in altermagnets. For a given energy $\omega$, the wavevector is spin-dependent and can be parameterized as $\bm{k}_{\sigma}(\omega,\theta) = k_{x,\sigma}(1,f_{\sigma})$  where $k_{x,\sigma}(\theta) = \sqrt{ (\omega+\mu)/[B(1+f_{\sigma}^2)/2 +\sigma \alpha_{A}f_{\sigma}]}$ and $f_{\sigma}(\theta)= -f_{-\sigma}(-\theta) = (B\tan\theta - \sigma \alpha_{A})/(B - \sigma\alpha_{A} \tan\theta)$, and the propagation angle is defined by $\tan\theta =y/x$~\cite{SM}. 
Here, the parameter $\sigma$ in the expressions (as opposed to the subscripts) takes values $\sigma = \pm 1$, corresponding to spin-up $(\uparrow)$ and spin-down $(\downarrow)$ states, respectively.
The $d_{xy}$-symmetry of the altermagnetic Fermi surface is reflected by the relation $M_x\bm{k}_\sigma(\omega,\theta)M_x^{-1}=\bm{k}_{-\sigma}(\omega,-\theta)$, where $M_x$ denotes mirror reflection about the $x$-axis. For a given displacement $\bm{r}$, the propagator is dominated by $\bm{k}_{\sigma}$ state whose group velocity $\bm{v}(\bm{k}_{\sigma})=\nabla _{\bm{k}_\sigma}\omega(\bm{k}_\sigma)$ is aligned with $\bm{r}$.

Due to the altermagnetic spin splitting, the propagators for opposite spins are modulated by distinct factors $\bm{k}_\sigma \cdot \bm{r}$ in Eq.~\eqref{eq2}. As a result, when the initial spin is oriented along the $x$ direction, the electron undergoes a distinctive spin precession [cf. Fig.~\ref{fig1}(b)], with the wave function given by
\begin{equation}\label{eq3}
    \psi(\bm{r};\omega) =
    - \frac{2\sqrt{2}\pi}{\sqrt{B^2-\alpha_{A}^2}}
    \begin{pmatrix}
        K_0(-i \bm{k}_{\uparrow}\cdot\bm{r}) \\
        K_0(-i \bm{k}_{\downarrow}\cdot\bm{r})
    \end{pmatrix}.
\end{equation}
The feature of altermagnetic spin precession becomes evident in the far-field limit $\bm{k}_\sigma\cdot\bm{r}\rightarrow\infty$, where the modified Bessel function exhibits the asymptotic behavior
$
        K_0(-i\bm{k}_{\sigma}\cdot\bm{r}) \rightarrow \sqrt{\frac{\pi}{2\bm{k}_{\sigma}\cdot\bm{r}}}e^{i(\bm{k}_{\sigma}\cdot\bm{r}+\frac{\pi}{4})}
$~\cite{arfken2011mathematical}. In this limit, the spatial spin distribution can be evaluated by the wave function as $(S_{x},S_y)=(\psi^\dag\sigma_x\psi,\psi^\dag\sigma_y\psi)=A(\bm{r})\left[\cos(\Delta \bm{k}\cdot \bm{r}),~ -\sin(\Delta \bm{k}\cdot \bm{r})\right]$, with $\Delta \bm{k}=(\Delta k_x, \Delta k_y)=\bm{k}_\uparrow - \bm{k}_\downarrow$ and $A=\left[(B^2-\alpha_A^2)\sqrt{(\bm{k}_\uparrow\cdot \bm{r})(\bm{k}_\downarrow\cdot \bm{r})}/(8\pi^3)\right]^{-1}$ the amplitude.
The distinctive spin precession is visualized in Fig.~\ref{fig1}(b). Specifically, as $\theta$ increases from $0$ to $\pi/4$, the spin precession rate gradually increases from zero (vanishing precession) to its maximum; and with further increasing $\theta$, the spin precession rate begins to decrease monotonically, until it reaches zero at $\theta=\pi/2$. Furthermore, the spin precession is mirror symmetric relative to the $x$ axis, \emph{i.e.}, the precessions for $\theta$ and $-\theta$ have the same rate but opposite directions.
As a result, the spatial spin distribution satisfies $S_y(x, y) = -S_y(x, -y)$, which can be verified in Fig.~\ref{fig1}(c).

In addition to the symmetric structure, the distinctive spin patterns are marked by the contours defined by $S_y(\bm{r})=0$, corresponding to the condition $\Delta \bm{k}\cdot\bm{r}=n\pi$, with $n$ an integer; see Fig.~\ref{fig1}(c). Notably, for $y\gg x$, we have $\Delta k_y \simeq 0$ and so the condition reduces to
\begin{equation}
x_n=n \mathcal{T}_x, \ \ \mathcal{T}_x=\frac{\lambda_F}{4}\sqrt{(B/\alpha_A)^2-1},
\label{Tx}
\end{equation}
where $\lambda_F=2\pi\sqrt{B/2\mu}$ is the Fermi wavelength without altermagnetism.
This result shows that the spin pattern exhibits periodic oscillations along the $x$ direction, with the spatial period $\mathcal{T}_x$ providing a direct measure of the altermagnetic spin splitting.

In typical transport experiments, electrons are injected from electrodes of finite width $W$ as shown in Fig.~\ref{fig1}(a), referred to as line injection. Specifically, we consider injection at $x=0$ over a finite region $y\in [-W/2, W/2]$, with spins polarized along the $x$ direction. The interference between different injection points is negligible, allowing the resulting spin textures to be interpreted as a simple superposition of the contributions from all points $\bm{r}'=(0,y')$ along the injection line, expressed as
\begin{equation}
     \begin{aligned}
     \tilde{S}_{x(y)}(\bm{r}) = \int_{-W/2}^{W/2} dy' S_{x(y)}(\bm{r}-\bm{r}').
     \end{aligned}
    \label{line}
\end{equation}
The spin distribution $\tilde{S}_y(\bm{r})$ for the line injection is shown in Fig.~\ref{fig1}(d). Similar to the case of point injection, the spatial profile of $\tilde{S}_y(\bm{r})$ remains odd with respect to the $x$-axis. Although the overall spin pattern for line injection differs significantly from that of point injection, it is noteworthy that the spin modulation along the $x$ direction near $y \to 0^\pm$ resembles that at $y = \pm W/2$ for point injection, as denoted by the solid/dashed arrows in Figs.~\ref{fig1}(c,d). To understand this, we consider the case $y > 0$ without loss of generality. Since $S_y(\bm{r};0,y')$ is an odd function of $y - y'$, the integral over the interval $[2y - W/2, W/2]$ in Eq.~\eqref{line} vanishes. Consequently, the remaining integral is over the interval $[-W/2, 2y - W/2]$, which reduces to the point injection at $(0,-W/2)$ as $y \to 0^+$. A similar argument holds for the case $y \to 0^-$ as well. More importantly, although the overall spin pattern for the line injection differs significantly from that of point injection, the spatial period given in Eq.~\eqref{Tx} remains preserved at the boundaries $y=\pm W/2$ for $x\ll W$~\cite{SM}, thereby facilitating its detection through transport measurements.
\begin{figure}[!htbp]
    \centering
    \includegraphics[width=0.48\textwidth]{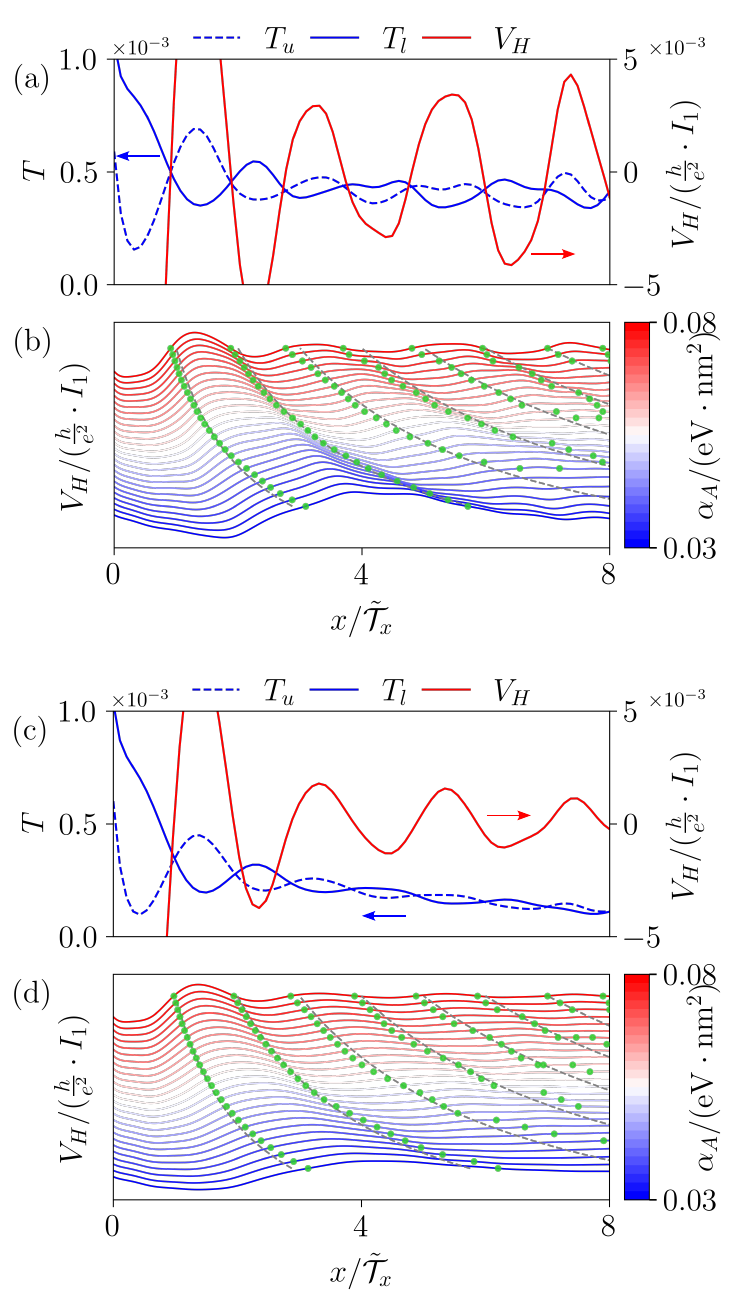}
    \caption{(a) Transmission probabilities from lead 1 to the upper (blue dashed line) and lower (blue solid line) probes and Hall voltage (red solid line) as a function of $x$ with $\alpha_{A} = 0.08\mathrm{~eV}\cdot \mathrm{nm}^2$.
(b) {A waterfall-like plot of} Hall voltage as a function of $x$ for different altermagnetic splitting $\alpha_{A}$, with its zeros marked by green points. The gray dashed lines correspond to $\tilde{x}_n(\alpha_A)$.
(c, d) Same plots including the dephasing effect with $\Gamma_{v}=3\mathrm{~meV}$.
The size of the scattering region is $50\mathrm{~nm}\times 50\mathrm{~nm}$, and the lattice constant is $a = 0.5\mathrm{~nm}$. The interface hopping is set to $t$ for leads 1 and 2, and to $0.1t$ for the transverse leads.
All other parameters are the same as those in Fig.~\ref{fig1}, which corresponds to the spatial period $\tilde{\mathcal{T}}_x = 4.67 \text{ nm}$.
}
    \label{fig2}
\end{figure}

\emph{Nonlocal spin-resolved transport.---}We consider spin-resolved transport measurements in the Hall bar device as illustrated in Fig.~\ref{fig1}(a). The setup consists of a square-shaped altermagnet connected to two longitudinal leads (leads 1 and 2), along with multiple transverse narrow leads designed to probe the spatially-resolved spin information. The transport properties are calculated based on the lattice model.
Specifically, the $d_{xy}$-type altermagnet can be mapped onto a 2D square lattice with lattice constant $a$ as~\cite{hallberg2025visualization}
\begin{equation}
    \begin{aligned}
     {H}_{\text{latt}} =& \epsilon\sum_{\textbf{i}}  {c}_{\textbf{i}}^\dagger {c}_{\textbf{i}}
      + t\sum_{ \langle \textbf{i},\textbf{j}\rangle} {c}_{\textbf{i}}^\dagger {c}_{\textbf{j}}
      + t_\alpha \sum_{ \langle\langle \textbf{i},\textbf{j}\rangle\rangle} \beta{c}_{\textbf{i}}^\dagger\sigma_z {c}_{\textbf{j}},
    \end{aligned}
    \label{latt}
\end{equation}
where ${c}_{\textbf{i}}=({c}_{\textbf{i}\uparrow},{c}_{\textbf{i}\downarrow})^{\text{T}}$ is the spinor fermionic operator at site $\textbf{i}$, the on-site potential is given by
$\epsilon = 4B/a^2 - \mu$, the nearest-neighbour hopping (denoted by $\langle\cdot\rangle$) is $t = -B/a^2$, and the next-nearest-neighbour hopping  (denoted by $\langle\langle\cdot\rangle\rangle$)  is $t_{\alpha} = \alpha_{A}/(4a^2)$, which accounts for the altermagnetic spin splitting. The factor $\beta=\pm$ corresponds to the hopping along the $(1,1)$ and $(1,-1)$ directions, respectively. The model parameters are designed such that Eq.~\eqref{latt} reduces to Eq.~\eqref{kp} in the long-wavelength limit. For all the leads, we adopt an effective model of normal electrons, given by $H_{\text{lead}}=\frac{B}{2}(k_x^2+k_y^2)-\mu+M\bm{\sigma}\cdot\hat{\bm{d}}$, which is also mapped onto the square lattice. In order to probe the spatial spin information, Zeeman splitting of strength $M$ is introduced to the leads with different normalized direction vectors $\hat{\bm{d}}$. Specifically, the Zeeman field in the injection lead (lead 1) is aligned along the $x$ direction, while those in all transverse detection leads are aligned along the $-y$-direction, both of which are in the ferromagnetic half-metal scenario. No Zeeman splitting is present in lead 2.

Next, we study the spin-resolved transport properties using the nonequilibrium Green's function method.
According to Landauer-B\"{u}ttiker formula~\cite{datta1997electronic}, the currents $I_p$ and voltages $V_{p,q}$ in different leads are related by the transmission functions $T_{pq}$ through
\begin{equation}
     \begin{aligned}
     I_{p} = \frac{e^2}{h} \sum_{q\neq p }(T_{qp}V_p - T_{pq}V_q).
     \end{aligned}
    \label{LB_formula}
\end{equation}
The transmission probability from lead $q$ to $p$ is calculated by $T_{pq}=\mathrm{Tr}\left[\Gamma_p G^R \Gamma_q G^A\right]$, where the retarded and advanced Green's functions are defined by $G^R=\left[G^A\right]^\dagger=\left[\omega-H_{\text{latt}} - \sum_p\Sigma_p^R\right]^{-1}$ and the linewidth function is given by $\Gamma_p = i\left(\Sigma^R_p - \Sigma^A_p\right)$. Both expressions contain the self-energies $\Sigma^R_p=\left[\Sigma^A_p\right]^\dag=t_p^2G^R_{s,p}$ due to the coupling between lead $p$ and scattering region with strength $t_p$, where $G^R_{s,p}$ is the surface Green's function of lead $p$~\cite{sancho1985highly}.
In our calculations, we consider current flowing between leads 1 and 2, with $I_{1}=-I_{2}$. The upper ($u$) and lower ($l$) leads, labeled by their positions $x$, serve as voltage probes that measure the transverse voltages $V_{u,l}(x)$ under the condition of vanishing current. The voltages at all leads, including $V_{1,2}$ and $V_{u,l}(x)$, are then solved by Eq.~\eqref{LB_formula}. To ensure experimental relevance, we choose model parameters that lie within a physically realistic regime for altermagnetic materials. Specifically, we set $B = 0.2~\mathrm{eV}\cdot\mathrm{nm}^2$ (corresponding to an effective mass $m^* = 0.38m_0$, with $m_0$ the bare electron mass), $\alpha_A = 0.08~\mathrm{eV}\cdot\mathrm{nm}^2$, and $\mu = 0.1~\mathrm{eV}$. These values are comparable to those reported for representative altermagnets, such as $\ce{KRu_4O_8}$ and $\ce{V_2OS}$~\cite{huang2025spin,vsmejkal2022emerging,guo2023spin}.

In Fig.~\ref{fig2}(a), we plot the transmission probabilities $T_{u}(x)$ and $T_{l}(x)$ from lead 1 to the upper and lower voltage probes as a function of their position $x$. The altermagnetic spin precession induces a periodic modulation of the spin overlap between the propagating electrons and the voltage probes, resulting in spatial oscillations of the transmission probabilities. By comparing with the spin textures near $y = \pm W/2$ in Fig.~\ref{fig1}(d), one can see that the modulation of transmission probabilities faithfully reveals the underlying spin precession. In particular, electrons reaching the upper and lower probes experience opposite spin precessions, such that their spin overlaps exhibit opposite spatial modulations. This explains the nearly $\pi$ phase shift between the oscillations of $T_{u}$ and $T_{l}$ in Figs.~\ref{fig2}(a,c).

These spatial modulations of the transmission probabilities are inherited by the Hall voltage, defined as $V_H(x) = V_{u}(x) - V_{l}(x)$; see Fig.~\ref{fig2}(a). Specifically, the potential at the voltage probe $p$ is given by
$V_p=\frac{\sum_{q\neq p}T_{pq}V_q}{\sum_{q\neq p}T_{pq}}$~\cite{datta1997electronic}.
In the present setup, this yields
$V_{u(l)}(x)\simeq\frac{T_{u(l)}V_1}{T_{u(l)}+T_{u(l)2}}$,
where $T_{u(l)2}$ denotes the transmission from lead 2 to the upper (lower) voltage probe. We have set $V_2=0$ without loss of generality and neglected the small transmissions between voltage probes at different positions $x$.
The Hall voltage vanishes at positions $\tilde{x}_n$ determined by $\tilde{S}_y(\tilde{x}_n,\pm W/2)=0$. At these points, the local spin overlaps with the upper and lower probes are identical, leading to $T_u=T_l$. Moreover, the symmetry of the system  ensures $T_{u2}=T_{l2}$ at $\tilde{x}_n$, which directly yields $V_H(\tilde{x}_n)=0$.
This correspondence is illustrated in Fig.~\ref{fig2}(b), where the Hall voltage modulations for different altermagnetic splittings are plotted, and the zero points (green dots) are compared with $\tilde{x}_n=n\tilde{\mathcal{T}}_x = 1.3 x_n$ (dashed lines), showing quantitative agreement. Here, the factor of $1.3$ arises from reflections at the upper and lower boundaries present in the transport calculations but absent in Fig.~\ref{fig1}(d).
Since $\mathcal{T}_x$ is determined by the strength $\alpha_A$ of altermagnetic spin splitting, our scheme provides a direct and experimentally accessible means for its measurement. 
The altermagnetic spin splitting spans a wide range, from $10~\mathrm{meV}$ to $1400~\mathrm{meV}$~\cite{vsmejkal2022emerging}, implying that the corresponding spin-precession period can vary from a few nanometers to several tens of nanometers. For a representative current of $I_1 = 1~\mathrm{\mu A}$, the corresponding Hall-voltage range in Figs.~\ref{fig2}(a,c) is $[-0.13, 0.13]~\mathrm{mV}$, which is well within experimental reach. The above analysis indicates that the predicted spin precession can be observed in intrinsic altermagnets such as KRu$_4$O$_8$ and V$_2$OS.

\emph{Dephasing and warping effects.---}The above calculations assume full phase coherence. In reality, however, dephasing and spin relaxation may occur due to inelastic scattering and interactions with magnetic impurities~\cite{wu2010spin, jiang2009topological}. Due to the altermagnetic coupling between spin and momentum, these two effects generally occur simultaneously.
To incorporate these effects, we employ B\"{u}ttiker’s approach, modeling dephasing by attaching virtual voltage probes~\cite{buttiker1986role,golizadeh2007nonequilibrium,zhou2022transport}. Each lattice site $\textbf{i}$ in the altermagnetic region is coupled to a probe represented by the self-energy $\Sigma_{\textbf{i}}^R=-i\Gamma_v/2$, where $\Gamma_v$ characterizes the dephasing strength, \textit{i.e.}, the probability for electrons to escape into the virtual probes and lose their phase information.
Mathematically, the self-energies of both real and virtual leads enter the Green’s function on an equal footing, \textit{i.e.}, $G^R=\left[\omega-H_{\text{latt}}-\sum_p\Sigma_p^R-\sum_{\textbf{i}}\Sigma_{\textbf{i}}^R\right]^{-1}$, from which the transmission matrix taking into account both real and virtual leads can be obtained. Using the same current configuration as in the case without dephasing, and imposing the condition of zero current through all voltage probes, the Hall voltages are solved by Eq.~\eqref{LB_formula}. In our calculations, we set $\Gamma_v = 3~\mathrm{meV}$, which corresponds to a coherent-to-total current ratio of 41\%.

The transmission probabilities $T_{u,l}$ and the Hall voltage $V_H$ as functions of $x$ are shown in Fig.~\ref{fig2}(c,d).
Due to the virtual voltage probes, both the direct transmissions $T_{u,l}$ and the Hall voltage $V_H$ decay with increasing $x$.
This decay occurs because electrons absorbed and reinjected by the virtual leads lose their phase and spin information.
Nevertheless, the oscillation amplitude of $V_H$ remains visible, and its spatial period $\tilde{x}_n$ is unaffected [cf. Fig.~\ref{fig2}(d)], demonstrating the robustness of the predicted effect against dephasing.
Moreover, the crystalline warping that may occur in real materials does not affect our main conclusions; see Supplemental Material~\cite{SM} for details.

\begin{figure}[!htbp]
    \centering
    \includegraphics[width=0.48\textwidth]{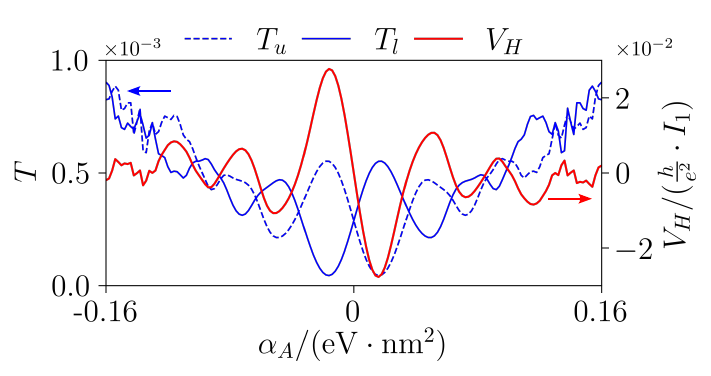}
    \caption{Transmission probabilities and Hall voltage as a function of $\alpha_{A}$ at $x=12.5\mathrm{~nm}$. All other parameters are the same as in Fig.~\ref{fig1}.
}
    \label{fig3}
\end{figure}

\emph{Altermagnetic spin transistor.---}Altermagnets have shown great potential for spintronic applications owing to their strong spin splitting, which is an order of magnitude larger than that induced by spin-orbit coupling~\cite{vsmejkal2022emerging}, thereby enabling spintronic devices with higher efficiency and smaller sizes. When altermagnetic spin splitting can be controlled electrically~\cite{wang2024electric}, the proposed device can function as a spin transistor~\cite{datta1990electronic}. We have seen that, for a fixed altermagnetic spin splitting $\alpha_A$, the spin precession gives rise to a location-dependent Hall voltage $V_H$, with a period determined by $\alpha_A$ through Eq.~\eqref{Tx}. Consequently, tuning $\alpha_A$ is expected to modulate both the spin precession and therefore $V_H$, thereby realizing a spin transistor. In Fig.~\ref{fig3}, we show the transmissions $T_{u,l}$ and $V_H$ as functions of $\alpha_A$, where only a single pair of Hall probes is involved. The results show that the voltage drop can be efficiently tuned, enabling clear switching with a high on/off ratio.

Physically, the external electric field is produced by dual gates ($V_{tg}$ and $V_{bg}$ in Fig.~\ref{fig1}(a)) placed on the top and bottom of the sample, which break certain crystalline symmetries and induce altermagnetic spin splitting. Taking the candidate $\ce{CaMnSi}$ as an example, the electric field breaks the combined parity-time-reversal symmetry while preserving $\left[\mathcal{T}||\mathcal{C}_{4z}\right]$ symmetry~\cite{wang2024electric}.
The altermagnetic splitting originates from the potential difference between the two nonmagnetic layers and can thus be tuned and even reversed.
For opposite spin splittings, the general relations $T_{u}(\alpha_{A}) = T_{l}(-\alpha_{A})$ and $V_{H}(\alpha_{A}) = -V_{H}(-\alpha_{A})$ hold, as confirmed in Fig.~\ref{fig3}.

It is instructive to highlight the differences between the altermagnetic spin transistor and the Datta-Das spin transistor~\cite{datta1990electronic}, which relies on gate-controlled spin-orbit coupling. Although both scenarios share a collinear spin structure, their distinct dimensionalities and momentum dependences give rise to qualitatively different spin dynamics and physical outcomes. In particular, Datta-Das spin precession manifests itself as oscillations in a longitudinal transport signal, whereas altermagnetic spin precession is revealed through oscillations in a transverse transport response. The larger spin splitting intrinsic to altermagnets enables high-frequency spin manipulation within a more compact device footprint, therefore enhancing robustness against dephasing. Furthermore, the broader range of candidate altermagnetic materials offers a promising route to overcome the long-standing spin-injection limitation due to the conductivity mismatch at interfaces~\cite{vzutic2004spintronics}.

\emph{Discussions.---}
The altermagnetic spin precession and transport properties are highly anisotropic and depend sensitively on the relative orientation between the applied bias and the crystallographic direction. In contrast to transport along the nodal direction, \emph{i.e.}, the $x$ direction for the $d_{xy}$ Fermi-surface configuration, transport along the anti-nodal direction yields a vanishing $V_H$; see the Appendix for details. This pronounced anisotropy further highlights the distinctive characteristics of altermagnetism.
Although we focus here on the $d$-wave altermagnet, similar analyses can be carried out for other symmetries~\cite{vsmejkal2022emerging,vsmejkal2022beyond}, where intriguing spin precession is likewise expected.
As this spin precession represents an intrinsic physical property of altermagnets, it has broad implications for their spintronic applications, such as generating complex spin textures, controlling magnetization dynamics, and manipulating domain walls and skyrmions~\cite{manchon2019current,obata2008current}.

\begin{acknowledgments}
We thank Gu Zhang for the helpful feedback on our manuscript.  
This work was supported by the National Natural Science Foundation of China (Nos. 12574051, 12222406, 12188101), the Natural Science Foundation of Jiangsu Province (Nos. BK20250008, BK20233001, BK20253009), the Fundamental Research Funds for the Central Universities (Nos. KG202501 and 2024300415), and the National Key Projects for Research and Development of China (No. 2022YFA1204701). This work has been supported by the New Cornerstone Science Foundation. 
\end{acknowledgments}

\emph{Note added.---}Recently, we became aware of a related work on altermagnetic spin transistors~\cite{zhu2025altermagnetoelectric}, which is based on multiferroic altermagnets and differs from the spin-precession-based mechanism considered in our work.

\section*{Data Availability}
The numerical data generated in this study have been deposited in the Zenodo database under accession code~\cite{liu_2026_18709554}.



\hspace*{\fill}

\ \ \ \ \ \ \ \ \ \ \ \ \ \ \ \ \ \ \ \ \ \ \ \textbf{End Matter}

\emph{Appendix.---}Here, we examine spin-resolved transport along the anti-nodal direction of the altermagnetic Fermi surface and compare the results with those presented in the main text. This situation can be equivalently analyzed by considering transport along the $x$ direction while adopting a $d_{x^2-y^2}$-type order, which is described by
\begin{equation}
H = \frac{B}{2}(k_x^2+k_y^2) + \frac{\alpha_A}{2}(k_x^2 - k_y^2)\sigma_z - \mu,
\end{equation}
where the parameter definitions follow those in Eq.~\eqref{kp}. Similar to Figs.~\ref{fig1}(c,d), we calculate the spatial distributions of the spin components $S_y$ and $\tilde{S}_y$.
\begin{figure}[!htbp]
    \centering
    \includegraphics[width=0.48\textwidth]{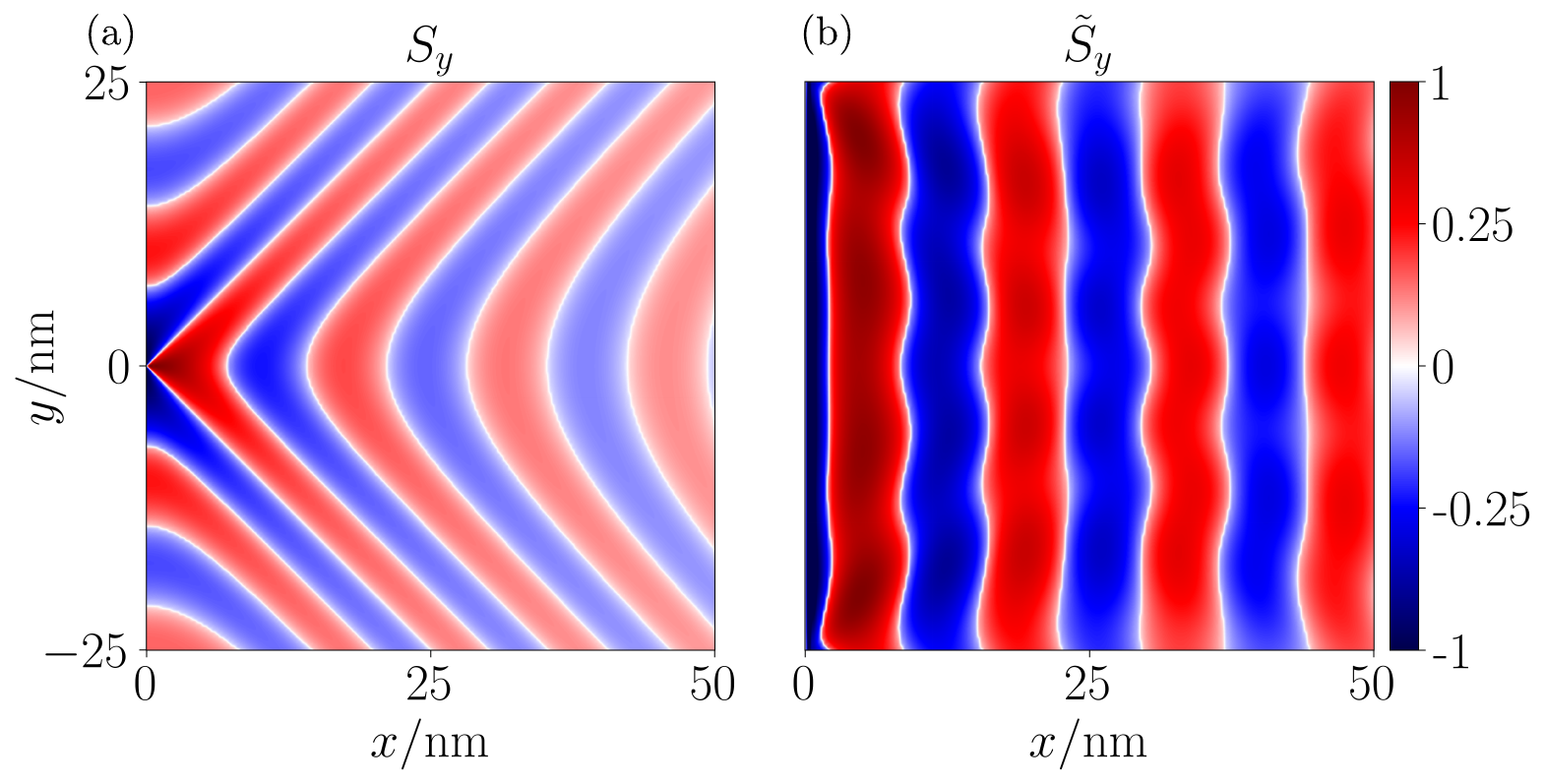}
    \caption{Spatial distribution of spin component for nonequilibrium propagating electrons under (a) point injection and (b) line injection, with the injected spins polarized along the $x$ direction. A square-root scale is applied to the colorbar to enhance the contrast around zero. All other parameters are the same as those in Fig.~\ref{fig1}.
}
    \label{fig4}
\end{figure}
As illustrated in Fig.~\ref{fig4}(a), the spatial distribution of $S_y$ under point injection is related to that in Fig.~\ref{fig1}(c) by a clockwise rotation of $\pi/4$, reflecting the transformation of the Fermi surface from the $d_{xy}$ type to the $d_{x^2-y^2}$ type. In contrast to Fig.~\ref{fig1}(c), the spin pattern here exhibits mirror symmetry with respect to the $x$ axis, consistent with the internal mirror symmetry of each spin band in the $d_{x^2-y^2}$ configuration.
Similarly, $\tilde{S}_y$ under line injection is also symmetric about the $x$ axis, resulting in a vanishing $V_H$. Interestingly, the pattern of $\tilde{S}_y$ also exhibits a spatial oscillation with a period
\begin{equation}
\mathcal{T}'_x= \frac{\lambda_F}{2}\frac{\sqrt{1-\alpha_A^2/B^2}}{\sqrt{1+\alpha_A/B}-\sqrt{1-\alpha_A/B}},
\end{equation}
which likewise encodes information about the altermagnetic spin splitting.

\onecolumngrid
\clearpage
\onecolumngrid 


\begin{center}
    {\large \textbf{Supplemental Material for: Altermagnetic Spin Precession and Spin Transistor}} \\
    \vspace{15pt}
    
    {Li-Shuo Liu,$^{1}$ Kai Shao,$^{1,3}$ Hai-Dong Li,$^{1}$ Xiangang Wan,$^{1}$ Wei Chen \orcidA{},$^{1,2,*}$ and D. Y. Xing$^{1}$} \\
    \vspace{8pt}
    
    {\small \it $^1$National Laboratory of Solid State Microstructures, School of Physics, \\ and Collaborative Innovation Center of Advanced Microstructures, Nanjing University, Nanjing 210093, China} \\
    \vspace{2pt}
    {\small \it $^2$Jiangsu Physical Science Research Center and Jiangsu Key Laboratory of Quantum \\ Information Science and Technology, Nanjing University, Nanjing 210093, China} \\
    \vspace{2pt}
    {\small \it $^3$School of Physics and Astronomy, Yunnan Key Laboratory for Quantum Information, \\ Yunnan University, Kunming 650091, People's Republic of China} \\
    
    \vspace{10pt}
    {\footnotesize $^*$Corresponding author: pchenweis@gmail.com}

    \vspace{25pt}
\end{center}
\setcounter{equation}{0}
\setcounter{figure}{0}
\setcounter{table}{0}
\setcounter{page}{1}
\setcounter{section}{0}

\renewcommand{\theequation}{S.\arabic{equation}}
\renewcommand{\thefigure}{S.\arabic{figure}}
\renewcommand{\thetable}{S.\arabic{table}}
\renewcommand{\thepage}{\arabic{page}}

\section{Derivation of Green's function}
In this section, we show that the Green's function can be expressed as the form of Eq.~(2) in the main text.
The effective Hamiltonian of 2D altermaget with $d_{xy}$ type order is given by
\begin{equation}
    H = \frac{B}{2}(k_x^2 + k_y^2) + \alpha_A k_x k_y\sigma_z - \mu.
\end{equation}
The retarded Green’s function is then written as
\begin{equation}
    g^R(\bm{r};\omega) = \int dk_xdk_y\frac{e^{ik_x x}e^{ik_yy}}{\omega+i\eta - H}.
\end{equation}
Since the Hamiltonian is diagonal in spin space, the Green’s function can be evaluated separately for each spin sector. For the spin-up component, we obtain
\begin{equation}
    g_{\uparrow}^R(\bm{r};\omega) = \int dk_xdk_y\frac{e^{ik_x x}e^{ik_yy}}{\omega+i\eta - \left[\frac{B}{2}(k_x^2 + k_y^2) + \alpha_A k_x k_y - \mu\right]}.
\end{equation}
First, we perform a $\pi/4$ rotation of the integration variables $k_x$ and $k_y$ to eliminate the coupling term $k_x k_y$, which gives
\begin{equation}
    g_{\uparrow}^R(\bm{r};\omega) = \int dk_xdk_y\frac{e^{ik_x (x+y)/\sqrt{2}}e^{ik_y (x-y)/\sqrt{2}}}{\omega+i\eta - \left[B_+k_x^2 + B_-k_y^2 - \mu\right]},
\end{equation}
where $B_{\pm} = (B\pm \alpha_A)/2$. Introducing the substitution $\sqrt{B_-/B_+}k_y \rightarrow k_y$ and switching to cylindrical coordinates, $k_x = k\cos\theta$ and $k_y = k\sin\theta$, we obtain
\begin{equation}
    g^R_{\uparrow}(\bm{r};\omega)=\sqrt{\frac{B_+}{B_-}}\int kdkd\theta \frac{e^{ik\left[\cos\theta\frac{x+y}{\sqrt{2}} + \sin\theta\sqrt{\frac{B_+}{B_-}}\frac{x-y}{\sqrt{2}}\right]}}{\omega + i\eta -(B_+ k^2 - \mu)}
    = \sqrt{\frac{B_+}{B_-}}\int_0^\infty k dk \frac{2\pi J_0(k\tilde{r}_{\uparrow})}{\omega + i\eta - (B_+k^2-\mu)},
\end{equation}
with $\tilde{r}_{\uparrow} = \sqrt{\left(\frac{x+y}{\sqrt{2}}\right)^2 + \left(\sqrt{\frac{B_+}{B_-}}\frac{x-y}{\sqrt{2}}\right)^2}$ and $J_0$ is the zeroth-order Bessel function of the first kind. Using the identity
\begin{equation}
    \int_0^\infty dx \frac{x J_0(ax)}{x^2+\beta^2} = K_0(a\beta), 
\end{equation}
where $K_0$ zeroth-order modified Bessel function of the second kind, we obtain
\begin{equation}\label{7}
    g^R_{\uparrow}(\bm{r};\omega)= -\frac{4\pi}{\sqrt{B^2- \alpha_{A}^2}}K_0(-i\tilde{k}_{\uparrow}\tilde{r}_{\uparrow}),
\end{equation}
with $\tilde{k}_{\uparrow} = \sqrt{\frac{\omega+\mu}{B_+}}$. 
The spin-down component of the Green’s function is obtained by substituting $\alpha_A \rightarrow -\alpha_A$, which gives
\begin{equation}\label{8}
    g_{\downarrow}^R(\bm{r};\omega) = -\frac{4\pi}{\sqrt{B^2- \alpha_{A}^2}}K_0(-i\tilde{k}_{\downarrow}\tilde{r}_{\downarrow}),
\end{equation}
with $\tilde{k}_{\downarrow} = \sqrt{\frac{\omega+\mu}{B_-}}$, $\tilde{r}_{\downarrow}=\sqrt{\left(\frac{x+y}{\sqrt{2}}\right)^2 + \left(\sqrt{\frac{B_-}{B_+}}\frac{x-y}{\sqrt{2}}\right)^2}$.

The physical picture of the propagators becomes clearer when they are recast into a different form. To this end, we analyze electron propagation in a semiclassical way. Consider an electron propagating from the origin to a position $\bm{r}$ with energy $\omega$, such that its velocity is aligned with the propagation direction. The corresponding equations for the wave vector are then
\begin{equation}
    \begin{aligned} 
     \frac{B}{2}(k_x^2+k_y^2) + \alpha_A k_x k_y -\mu = \omega,\\
     \frac{v_y}{v_x}=\frac{Bk_y+\alpha_A k_x}{Bk_x + \alpha_A k_y} = \frac{y}{x} \equiv  \tan\theta.
    \end{aligned} 
\end{equation}
Solving the equations yields
\begin{equation}
    \begin{aligned} 
     k_{x,\uparrow}(\omega,\theta) = \sqrt{\frac{\omega+\mu}{B\left[1+f_{\uparrow}(\theta)^2\right]/2+\alpha_A f_{\uparrow}(\theta)}},\ \ 
     k_{y,\uparrow}(\omega,\theta) = f_{\uparrow}(\theta) k_{x,\uparrow}(\omega,\theta), \ \ 
     f_{\uparrow}(\theta) = \frac{B\tan\theta-\alpha_A}{B-\alpha_A\tan\theta}.
    \end{aligned} 
\end{equation}
The wave vector for spin-down electrons is obtained by replacing $\alpha_A \rightarrow -\alpha_A$. The signs of $k_x$ and $k_y$ are chosen such that $\bm{v} \parallel \bm{r}$, ensuring that the group velocity is aligned with the propagation direction.

Using the wave vectors defined above, it is straightforward to show that
\begin{equation}
    \begin{aligned} 
     \tilde{k}_{\sigma}\tilde{r}_{\sigma} &= \bm{k}_{\sigma}(\omega,\theta)\cdot\bm{r}.
    \end{aligned} 
\end{equation}
The propagator then takes the form
\begin{equation}
    g^R_{\sigma,\sigma'}(\bm{r};\omega) = - \frac{4\pi\delta_{\sigma,\sigma'}}{\sqrt{B^2- \alpha_{A}^2}}K_0(-i {\bm{k}}_{\sigma}\cdot\bm{r}),
\end{equation}
which corresponds to Eq.~(2) in the main text. This expression has a transparent physical interpretation: the phase accumulation during propagation is governed by the wave vector whose associated group velocity is aligned with the displacement. Consequently, the phase modulation, and thus the spin precession, can be analyzed within a semiclassical framework.
\begin{figure}[htbp]
    \centering
    \includegraphics[width=0.8\textwidth]{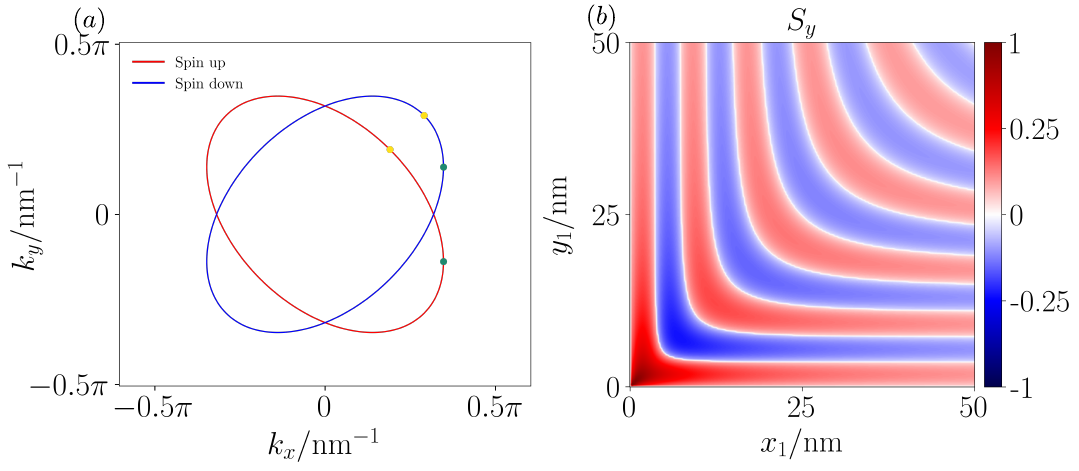}
    \caption{(a) The Fermi surface of altermagnet with wavevectors whose group velocity along $\theta = 0,\pi/2$ are marked by green and yellow dots, respectively.
    (b) Spatial distribution of $S_y(\bm{r}_1)$ for nonequilibrium propagating electrons. A square-root scale is applied to the colorbar in order to emphasize the contrast around zero.}
    \label{fig_S1}
\end{figure}

To gain further insight into the above formula, we consider two representative directions, $\theta = 0$ and $\theta = \pi/4$. As shown in Fig.~\ref{fig_S1}(a), when the group velocity is along $\theta = 0$, the corresponding wavevectors (green dots) are symmetric about the $x$-axis, yielding $(\bm{k}_{\uparrow} - \bm{k}_{\downarrow}) \perp \bm{r}$ and thus no spin precession along this direction. In contrast, for $\theta = \pi/4$, the wavevectors (yellow dots) are aligned with $\bm{r}$, resulting in the maximal precession frequency, $(\bm{k}_{\uparrow} - \bm{k}_{\downarrow}) \cdot \bm{r}/|\bm{r}|$.

\section{Period of spin modulation}
In this section, we show that the spatial period of the spin modulation corresponding to the point injection can be expressed in the form of Eq.~(4) in the main text. Specifically, the spatial distribution of $S_y$ is given by
\begin{equation}
    S_y(\bm{r}) = -A(\bm{r})\sin(\Delta \bm{k}\cdot \bm{r}),\ \ 
    \Delta \bm{k} = (\Delta k_x, \Delta k_y) = \bm{k}_{\uparrow} - \bm{k}_{\downarrow},\ \ 
    A(\bm{r})=\left[ (B^2 - \alpha_A^2) \sqrt{(\bm{k}_{\uparrow}\cdot \bm{r})(\bm{k}_{\downarrow}\cdot \bm{r})}/(8\pi^3)\right]^{-1}.
\end{equation}
The contours defined by $S_y(\bm{r}) =0$ satisfy
\begin{equation}
    \left[\bm{k}_{\uparrow}(\omega,\theta) - \bm{k}_{\downarrow}(\omega,\theta) \right]\cdot\bm{r} = n\pi,\quad n = 0, \pm 1, \pm 2,\cdots, 
    \label{eq14}
\end{equation}
which yields a family of real-space curves $y = y(x)$, each corresponding to an integer number of spin-precession periods. 
A general parametric expression for these curves is difficult to obtain analytically. In the limit $\theta \rightarrow \frac{\pi}{2}^-$, however, one finds $\Delta k_y(\omega, \frac{\pi}{2}^-) \simeq 0$, and moreover, 
$\lim_{\theta\rightarrow \frac{\pi}{2}^-} \Delta k_y(\omega, \frac{\pi}{2}^-) y = 0$. The latter can be verified by introducing the substitution $\tan\theta \rightarrow 1/t$, under which $\Delta k_y y$ becomes
\begin{equation}
    \begin{aligned}
    \Delta k_y y
    = & \left[(B-\alpha_A t)\sqrt{\frac{\omega + \mu}{(B^2-\alpha_A^2)(\frac{B}{2}t^2 -\alpha_A t + \frac{B}{2})}}
    - (B+\alpha_A t)\sqrt{\frac{\omega + \mu}{(B^2-\alpha_A^2)(\frac{B}{2}t^2 +\alpha_A t + \frac{B}{2})}}\right]\frac{1}{t}x\\
    \simeq&\left[-2\sqrt{\frac{2(\omega+\mu) B}{B^2 - \alpha_A^2}}\left(\frac{\alpha_A}{B} - \left(\frac{\alpha_A}{B}\right)^3\right) t^2 + O(t^3)\right]x.\\
    \end{aligned}
\end{equation}
As $\theta \rightarrow \frac{\pi}{2}^-$ (i.e., $t\rightarrow 0$), one then obtains $\Delta k_y y\rightarrow0$. This implies that for $y\gg x$, the spin modulation is entirely governed by variations in $x$, as evidenced by the vertical stripe patterns in Fig.~\ref{fig_S1}(b).

Therefore, the condition~\eqref{eq14} reduces to 
\begin{equation}
    \Delta k_x(\omega,\pi/2) x_n=n\pi,
    \label{period}
\end{equation}
which yields
\begin{equation}
    x_n = \frac{n\pi}{\Delta k_x(\omega,\pi/2)} = \frac{n\pi}{2k_{x,\uparrow}(\omega,\pi/2)} = \frac{n\pi}{2\alpha_A}\sqrt{\frac{B(B^2-\alpha_A^2)}{2(\omega + \mu)}}.
\end{equation}
The same results hold for the limit $\theta\rightarrow -\frac{\pi}{2}^+$ as well. For electrons at the Fermi energy ($\omega = 0$), this expression reduces to
\begin{equation}
    x_n = n\mathcal{T}_x, \quad \mathcal{T}_x = \frac{\lambda_F}{4}\sqrt{(B/\alpha_A)^2-1},
    \label{Tx_S}
\end{equation}
where $\lambda_F = 2\pi\sqrt{B/2\mu}$ is the Fermi wavelength defined without altermagnetism. This is Eq.~(4) in the main text.
This result shows that the spin pattern exhibits periodic oscillations along the $x$ direction for $y\gg x$.

Moreover, this spatial period persists in the line-injection regime. As shown in the main text, the spin density at $\bm{r} = (x,y)$ under line injection is given by
\begin{equation}
    \tilde{S}_{y}(\bm{r}) = \int_{-W/2}^{W/2} dy' S_{y}(\bm{r}-\bm{r}'), 
    \label{eq19}
\end{equation}
with $\bm{r}'=(0,y')$ denoting the position along the injection line.
Making the variable substitution  $\bm{r}_1 \equiv (x_1, y_1) = \bm{r} - \bm{r}' = (x, y-y')$ and focusing on the spin modulation along the horizontal line $y=W/2$, where measurements are performed, one obtains
\begin{equation}
    \tilde{S}_{y}(x,W/2) = \int_{0}^{W} dy_1 S_{y}(x, y_1).
\end{equation}
It is noteworthy that the dominant wave vectors $\bm{k}_{\uparrow,\downarrow}$ possess group velocities aligned with the direction of $\bm{r}_1$.
From the spin pattern $S_y(\bm{r}_1)$ illustrated in Fig.~\ref{fig_S1}(b), one can see that $\Delta k_y y_1 \simeq 0$ for $|y_1/x_1| \gtrsim  1 $, and therefore $S_y(\bm{r}_1)$ in this region can be approximated as 
\begin{equation}
    S_y(\bm{r}_1) \simeq A(\bm{r}_1)\sin(\Delta k_x x_1).
\end{equation}
Then the integral can be divided into two parts as
\begin{equation}
    \tilde{S}_{y}(x, W/2) \simeq \int_{0}^{x} dy_1 S_{y}(x, y_1) + \int_{x}^{W} dy_1 A(\bm{r}_1)\sin(\Delta k_{x} x)
    \simeq \int_{x}^{W} dy_1 A(\bm{r}_1)\sin(\Delta k_{x} x),
\end{equation}
where the second approximation is valid in the regime of interest with $x \ll W$. Meanwhile, $\Delta k_x(0,\theta)\simeq \Delta k_x(0,\pi/2)$ holds throughout most of the integration interval, allowing us to approximate further
\begin{equation}
        \tilde{S}_{y}(x,W/2) \simeq \sin\left[\Delta k_{x}(0,\pi/2) x\right] \int_{x}^{W} dy_1 A(\bm{r}_1).
\end{equation}
Consequently, the condition for the zeros coincides with Eq.~\eqref{period}, demonstrating the correspondence between point and line injections.

\section{Warping effect}
\begin{figure}[htbp]
    \centering
    \includegraphics[width=0.8\textwidth]{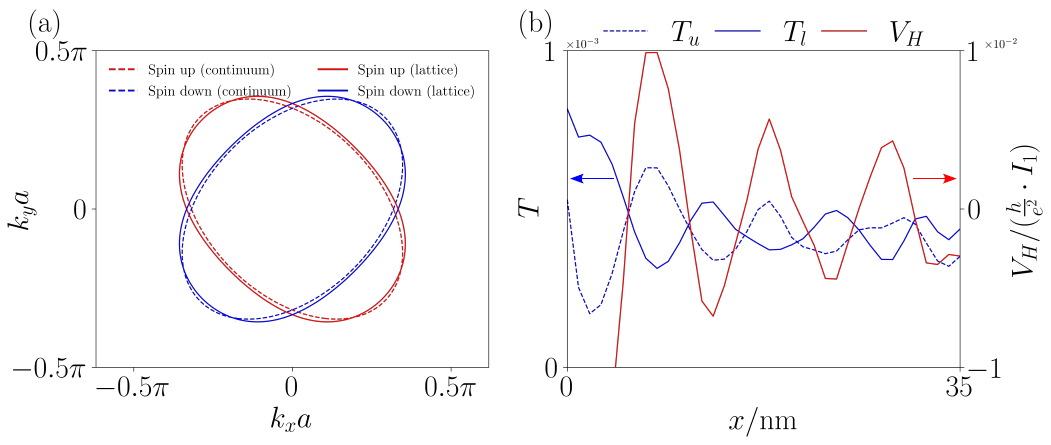}
    \caption{(a) Fermi surface of the lattice model including warping effects (solid lines) compared with the continuum-model Fermi surface (dashed lines). 
    (b) Transmission probabilities from the source to the upper (dashed blue) and lower (solid blue) probes as a function of $x$; the red curve shows the corresponding Hall voltage. The lattice constant is set to $a=1~\mathrm{nm}$ to simulate the warping effect, while the other parameters are the same as those in the main text.}
    \label{Warp_effect}
\end{figure}

In this section, we investigate the effect of crystalline warping on the spin-resolved transport properties of altermagnets, which may arise in real materials. 
To simulate this effect, we adopt a larger virtual lattice constant $a$, thereby intentionally moving the calculation away from the long-wavelength regime. This introduces deviations from the original parabolic band dispersion, resulting in a pronounced warping of the Fermi surface, as shown in Fig.~S2(a). The corresponding spin-resolved transport results, calculated using the same method, are presented in Fig.~\ref{Warp_effect}(b). The main features remain the same as those in Fig.~2(a) of the main text: the nearly $\pi$ phase shift between the oscillations of $T_{u}$ and $T_{l}$ is clearly preserved, and the Hall-voltage oscillations remain visible. Since $|k_{x,\uparrow}(\omega,\frac{\pi}{2}^-)|$ is slightly smaller than in the continuum model, the period of the Hall voltage increases accordingly, consistent with Eq.~\eqref{Tx_S}.

It should be emphasized that introducing warping effects by increasing the virtual lattice constant $a$ is purely a numerical strategy, and does not correspond to any real lattice distortion of an actual material. It simulates a pristine material with fixed physical parameters and a prescribed degree of warping.
This numerical procedure differs fundamentally from warping induced by real lattice distortions in materials, where all relevant physical parameters--such as $B$ and $\alpha_A$--would generally be modified in a correlated and material-dependent manner.

\end{document}